\documentclass[letterpaper, 10 pt, conference]{ieeeconf}  

\IEEEoverridecommandlockouts                              

\usepackage{graphics} 
\usepackage{epsfig} 
\usepackage{mathptmx} 
\usepackage{times} 
\usepackage{amsmath} 
\usepackage{amssymb}  

\usepackage{amsfonts}
\usepackage[T1]{fontenc}
\DeclareMathAlphabet{\mathpzc}{T1}{pzc}{m}{it}
\usepackage{comment}

\usepackage{url}

\title{\LARGE \bf
Physics-based Mesh Deformation with Haptic Feedback and Material Anisotropy
}

\author{Avirup Mandal$^{1}$, Parag Chaudhuri$^{2}$ and Subhasis Chaudhuri$^{1}$
\thanks{$^{1}$A. Mandal and S. Chaudhuri are with the Department of Electrical Engineering, Indian Institute of Technology Bombay, India.
        {\tt\small avirupmandal@ee.iitb.ac.in, sc@ee.iitb.ac.in}}%
\thanks{$^{2}$P. Chaudhuri is with the Department of Computer Science Engineering, Indian Institute of Technology Bombay, India.
        {\tt\small paragc@cse.iitb.ac.in}}%
}

\begin{document}

\maketitle
\thispagestyle{empty}
\pagestyle{empty}

\begin{abstract}
We present a physics-based framework to simulate porous, deformable materials and interactive tools with haptic feedback that can reshape it. In order to allow the material to be moulded non-homogeneously, we propose an algorithm to change the material properties of the object depending on its water content. We present a multi-resolution, multi-timescale simulation framework to enable stable visual and haptic feedback at interactive rates. We test our model for physical consistency, accuracy, interactivity and appeal through a user study and quantitative performance evaluation.
\end{abstract}
\section{Introduction}\label{sec:introduction}

Traditionally simulated virtual shape editing tools offer visual rendering of the object but entirely miss the tactile aspect of it. Moreover, many of these tools edit shapes in a purely geometric approach and thus, are not physically accurate. In this paper, we present a multi-resolution, multi-timescale framework for simulating stable, interactive, physics-based deformation with faithful tactile feedback. 

In real world an artist deforms a lump a clay to sculpt it into a model. Adding water to the clay makes it malleable, which in turn helps to reshape parts of the same model differently. Our method allows a user to perform these operations virtually while receiving the appropriate haptic feedback for the same. The major contributions of the work presented in this paper are as follows.
\begin{itemize}
    \item Develop multi-resolution, multi-timescale framework for stable, interactive, physics-based haptic and visual simulation.
    \item Develop physically valid model for deformable, porous soft objects represented as volumetric meshes.
    \item Modeling of objects with anisotropic elasticity when parts of the material are made wet. This allows the user to deform the object differently in different parts while applying the same force.
\end{itemize}
The rest of the paper is organized as follows. After presenting a discussion on the related works, we detail our simulation of variable elasticity produced by material wetting. Next, we present our haptic rendering solution for faithful tactile feedback. We then tie all the pieces of our framework for multi-resolution, multi-timescale simulation. Finally, we present qualitative, quantitative and user study results generated using our framework.
\section{Related Work}\label{related_work}
In this section we review the methods present in literature that are closely related to our work. One of the most popular approaches to model deformable objects is the Finite Element Method (FEM). O'Brien~\cite{ductile_frac}, M\"uller et al.~\cite{deform_muller} used FEM on tetrahedral meshes with linear elasticity to model deformable objects including plasticity and fracture. Non-linear elasticity with large plastic flow is rendered in more recent works by Bargteil et al.~\cite{deform_bargteil}, Irving et al.~\cite{deform_irving}.

Fluid flow and material wetting is a well studied subject in material physics~\cite{porous_bear}~\cite{porous_adler}. In computer graphics, work by Patkar et al.~\cite{wet_patkar} offers a geometrically modeled solution to absorption, diffusion and dripping of water in porous materials. In more recent work~\cite{wet_fei}, wetting of different kinds of clothes is explored. The change of material properties due to fluid absorption is well investigated in material science~\cite{elas_young}~\cite{elas_schraad}. Here the authors hypothesize a mathematical relationship of object elasticity with fluid content and finally verify their hypothesis with empirical results.

Zilles and Salisbury~\cite{god_obj_zilla} present God Object based rendering where a god object is constrained to stay on the surface of the mesh object while a haptic proxy penetrates into the object and the difference of their acceleration generates haptic feedback. This is improved by Ortega et al.~\cite{god_obj_ortega} by extending it to all six degrees of freedom. Another broad category of haptic force rendering is penalty based rendering~\cite{dis_pen_1}~ \cite{dis_pen_2}~\cite{dis_pen_3}. Here~\cite{ccd_hap_forc}, the colliding objects penetrate each other and force feedback is rendered depending on the depth of penetration. Discrete penalty based rendering suffers from discontinuous and jerky force feedback when the contact stiffness is high. These problems are circumvented using continuous collision detection~\cite{ccd_tang2}~\cite{ccd_hap_forc}. Even though constrained based methods are slightly more robust against the pop-through effect of proxy, we opted for the continuous penalty based method for our haptic feedback as it produces smoother force feedback~\cite{ccd_hap_forc}.

Notable works in virtual shape editing include methods presented in~\cite{vir_sculpt_1}~\cite{vir_sculpt_4} which build a rigid model using small cubic grid based field. These methods do not emulate physically accurate material behaviour and time consuming. In the works~\cite{vir_sculpt_2}~\cite{vir_sculpt_3}, the authors present frameworks that deform polygonal mesh based on mass-spring based models in a strictly geometric way. The key drawback of all these existing works is that none of them preserve physical plausibility. The need for physically realistic virtual sculpting has been explored recently by De Goes et al.~\cite{vir_sculpt_DeGoes}. Using Kelvinlets (fundamental solutions of linear elasticity for singular loads), they render the accurate mesh deformation in real-time, but their work lacks the aspects of haptic feedback. Moreover, the haptic feedback provided in all these works are based on discrete collision handling which suffers from jitters. We use continuous collision based smooth haptic feedback to tackle this problem.

We present a framework to efficiently reshape meshes in a physically realistic manner with smooth haptic feedback. Additionally, real-life objects have anisotropic elasticity that can be modelled implicitly using wetting in our framework.
\section{Modeling Deformable Porous Objects}\label{sec:wetting_mesh}
In this section we will briefly describe the modelling of deformable object using Cauchy's linear strain model and then discuss how we model the change of elasticity due to wetting of the object.

\subsection{Deformed Object Model}
We use a standard finite element discretization to solve the governing differential equations of a deforming object~\cite{deform_muller}~\cite{anime_erleben}. 
\begin{equation}\label{EQ:displace}
    \mathbf{u}(\mathbf{\zeta}, t) = \sum_{i=1}^{n_v} \mathbf{N}_i(\mathbf{\zeta})\mathbf{u}_i(t), \;\;\;\;\; \forall \mathbf{\zeta} \in \Theta
\end{equation}
where $\mathbf{N}(\mathbf{\zeta})$ and $\mathbf{u}_i(t)$ represents the shape function and displacement vector at the node $i$ respectively.
\begin{figure}
    \centering
    \includegraphics[width = 0.9\columnwidth]{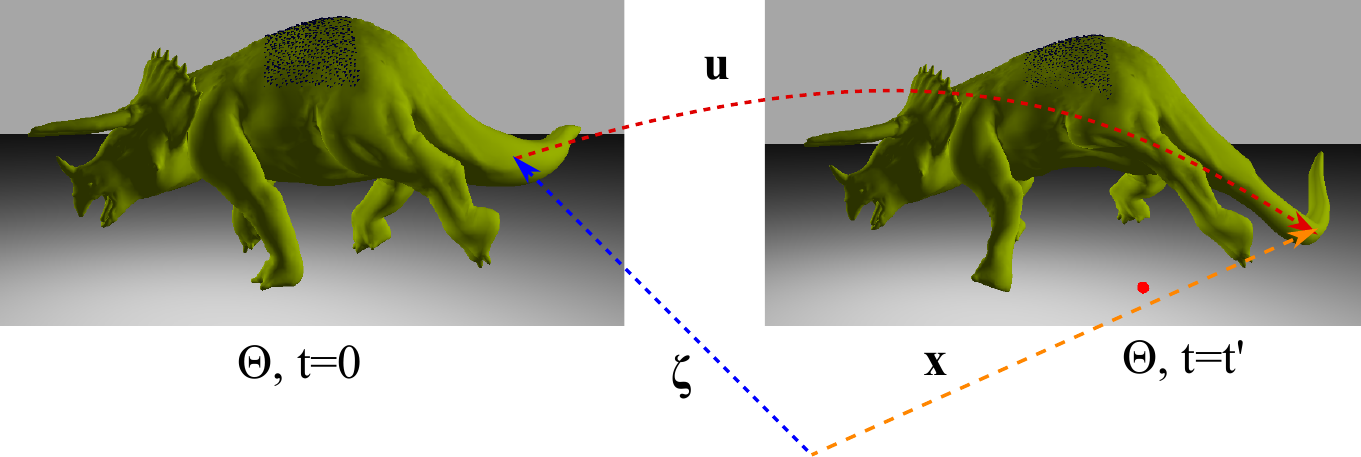}
    \caption{Material (left) coordinate to world (right) coordinate}
    \label{fig:deform}
\end{figure}
The system dynamics of deformed object can then be written in Lagrange's form as 
\begin{equation}\label{EQ:dynamic}
    \mathbf{M} \ddot{\overline{\mathbf{u}}} + \mathbf{f}^{int} = \mathbf{f}^{ext}, \; \;  \overline{\mathbf{u}} = \left(\mathbf{u}_{1}^{T} \hdots \mathbf{u}_{n_v}^{T} \right)^T
\end{equation}
where $\mathbf{M}$, $\mathbf{f}^{ext}$, $\mathbf{f}^{int}$ are respectively mass matrix, external and internal element force vector of the full system. Plastic flow is also enforced in this model when the strain exceeds a yield threshold as presented in the work by M\"{u}ller et al.~\cite{deform_muller}.

\subsection{Wetting Porous Object}
For wetting the material, we follow the method proposed in~\cite{wet_patkar} barring the dripping part of algorithm that reduces the fluid content in the object. Once collision occurs between the wetting tool in our framework and boundary of tetrahedral mesh, the fluid content in the tetrahedrons in contact with tool increases in incremental steps till the saturation value become one. The saturation of a tetrahedron is defined as $\displaystyle S_w = m^w/V^e$. Here $m^w$ is the mass of water absorbed and $V^e$ is the volume of tetrahedron. After the absorption of fluid, diffusion happens between any two neighbouring tetrahedra depending on the saturation gradient between them. Instead of dripping, a drying tool is provided whose action is complementary to the wetting one.

\subsection{Variable Elasticity}
In order to formulate a relationship between fluid content and elasticity of the material we followed the line of thought presented in~\cite{elas_young}. To the best of our knowledge, our framework is the first use of this method in an interactive, real-time setting. The Voigt upper bound on the elasticity tensor of a solid-fluid mixture, is given by 
\begin{equation}\label{EQ:elastic_upper}
    \mathbf{C}^V = \left(1 - \phi \right)\mathbf{C}_s + \phi \mathbf{C}_w
\end{equation}
where $\mathbf{C}^V$, $\mathbf{C}_s$ and $\mathbf{C}_w$ are the elasticity tensor of mixture, solid and fluid respectively. The quantity $\phi$ denotes the fluid volume fraction in solid. The Reuss lower bound on the compliance tensor of a solid-fluid mixture with any kind of solid and fluid, is given by  
\begin{equation}\label{EQ:compliance_lower}
    \mathbf{S}^R = \left(1 - \phi \right)\mathbf{S}_s + \phi \mathbf{S}_w
\end{equation}
where $\mathbf{S}^R$, $\mathbf{S}_s$ and $\mathbf{S}_w$ are the compliance tensor of mixture, solid and fluid respectively. The Voigt and Reuss bounds together give the bound on the effective elasticity tensor as
\begin{equation}\label{EQ:elas_upper_lower}
    \left(\mathbf{S^R}\right)^{-1} \leq \mathbf{C}^{eff} \leq \left(\mathbf{C}^V\right)
\end{equation}
Putting everything together, the effective compliance tensor for the solid-fluid mixture with any kind of solid and fluid, is given by 
\begin{equation}\label{EQ:compliance_effc}
    \mathbf{S}^{eff} = \left[\mathbf{1} + \phi\left(\mathbf{Q}^{I}-\mathbf{P}^{I}\right)^{-1}\right]\mathbf{S}^{M}
\end{equation}
where $\mathbf{Q}^{I} \equiv \left(\mathbf{C}^M - \mathbf{C}^I \right)^{-1} \mathbf{C}^M$, $\mathbf{P}^{I}$ is the Eshelby tensor~\cite{Eshelby_inclusion}, $\mathbf{S}^{M}$ is the matrix compliance tensor, $\mathbf{C}^{M}$ is the matrix elasticity tensor  and $\mathbf{C}^{I}$ is the inclusion elasticity tensor. For our framework, we assume $\mathbf{C}^M = \mathbf{C}_s$, $\mathbf{S}^M = \mathbf{S}_s$ and $\mathbf{C}^I = \mathbf{C}_w$. Using this formulation we determine the effective elastic tensor of a solid-fluid mixture system. We always use water for fluid in our framework. As the change of elasticity is dependent on the fraction of water content in the tetrahedral element and water content is dependent on the gradient of saturation, there is never any abrupt change of elasticity in the model, thus maintaining the stability of the system. 
\section{Haptic Rendering}\label{haptic_render}
The model developed in the previous section for deformation of an object made of a  porous material is used in conjunction with haptic rendering. The haptic interaction process with an object consists of the following components: 
\begin{itemize}
    \item Continuous Collision Detection (CCD) between the haptic proxy and tetrahedral simulation mesh.
    \item Continuous penalty based haptic rendering while deforming the mesh.
\end{itemize}
Haptic force feedback are performed on volumetric simulation mesh with tetrahedral elements. But to improve the quality of visually rendering, we transfer the deformation to a higher resolution visualization surface mesh. This is explained in Section~\ref{sec:multires}.

\begin{figure}[h!tb]
    \centering
    \includegraphics[width=0.9\columnwidth]{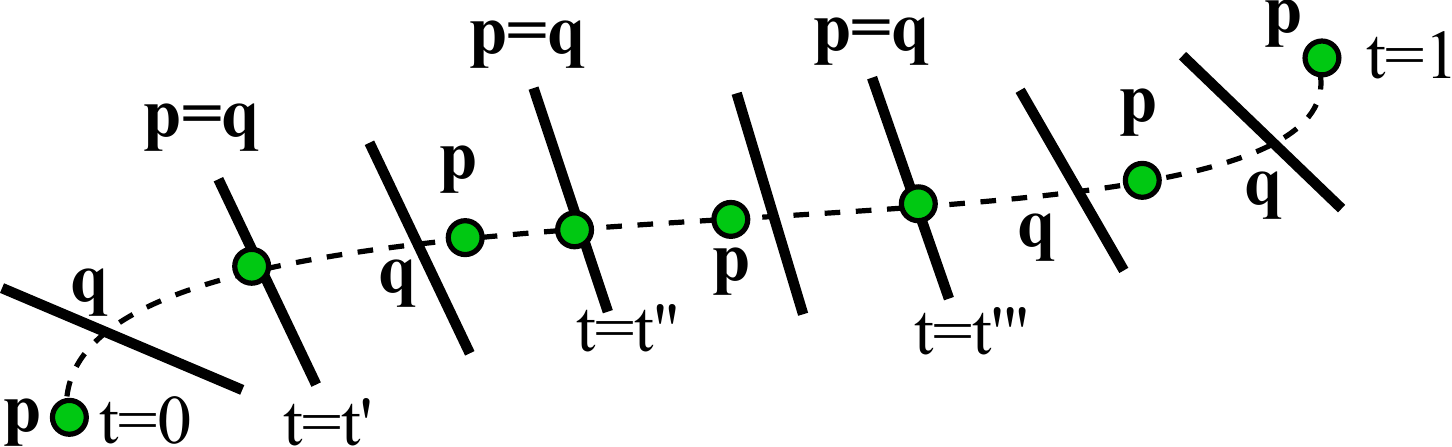}
    \caption{Three contact times are $t^{'},t^{''},t^{'''}$. Penetration intervals are $[t^{'},t^{''}]$ and $[t^{''}, 1]$. Time step $\displaystyle \Delta t$ is normalized to $\displaystyle [0, 1]$.}
    \label{fig:continuous_pen}
\end{figure}
\subsection{Continuous Collision: Vertex-Face and Edge-Edge}
We calculate continuous collision between the outer boundary of the tetrahedral mesh and haptic proxy, both of which consist of triangular face elements. We resolve vertex-face and edge-edge collisions that arise when two triangular face elements collide. In order to detect continuous collision, we begin by interpolating the positions of each primitive i.e., vertex, edge and face in the simulation time step, $\Delta t$, normalized to $[0,1]$. Then a $3^{rd}$ order equation in $t$ is solved to find out the number of collisions that occur between vertex-face or edge-edge interactions during that simulation time step. To detect the collisions fast, we used an Axis Aligned Bounding Box for each of these primitives and also followed a non-penetration filter based technique~\cite{ccd_tang1}.

\subsection{Haptic Rendering: Deforming a Mesh}\label{sec:haptics-deform}
We classify the deformation of a mesh in two categories: $(1)$ push deformation of mesh and $(2)$ pull deformation of mesh, both of which follow the same principle except the applied force direction which is inward for push and outward for pull. We detect collision using CCD in each time step, $\displaystyle \Delta t$, and then integrate over those particular time intervals when penetration depth between collided primitives is positive, implying that they are in a colliding state (see Figure~\ref{fig:continuous_pen}).

\subsubsection{Vertex-Face penalty force}
If a collision occurs between any vertex of the haptic proxy and the triangular mesh boundary of the object or vice-versa, then we calculate a penalty force~\cite{ccd_tang2} as
\begin{equation}\label{EQ:pen_forc}
    \mathbf{I}_p^{VF} = k_{vf}\sum_{i=0}^{i<N} \int_{t_a^i}^{t_b^i}\mathbf{n}_t^T\left(\mathbf{p}_t-\mathbf{q}_t\right)\mathbf{n}_tdt
\end{equation}
Here $k_{vf}$ is a scalar stiffness constant. Time intervals $[t_a^i,t_b^i]\in[0,1]$ are called penetration time intervals. These are defined as time while the vertex is inside the object mesh and $i$ is the number of penetration time intervals (See Figure~\ref{fig:continuous_pen}). Moreover $\mathbf{n}_t$, $\mathbf{p}_t$ and $\mathbf{q}_t$ denote contact normal, position of vertex and contact point on boundary mesh respectively during $\Delta t$. The collision point on the triangular mesh can be expressed using barycentric coordinates of three vertices of the mesh as $\displaystyle \mathbf{q}_t = w_a\mathbf{a}_t+w_b\mathbf{b}_t+w_c\mathbf{c}_t$. Once we get the penalty force $\mathbf{I}_p^{VF}$ we apply it to the object mesh $\displaystyle \left(-\mathbf{I}_p^{VF} \text{for pulling} \right)$ and also apply a reaction force of the same magnitude but opposite direction to the proxy.
 
\subsubsection{Edge-Edge penalty force}
Similar to vertex-face penalty force, we calculate penalty force $\displaystyle \mathbf{I}_p^{EE}$ due to edge-edge collision also. If a collision occurs between edge of the haptic proxy mesh and edge of the simulation mesh boundary of the object, then the penalty force is calculated as
\begin{equation}\label{EQ:pen_forc_ee}
    \mathbf{I}_p^{EE} = k_{ee}\sum_{i=0}^{i<N} \int_{t_a^i}^{t_b^i}\mathbf{n}_{E_t}^T\left(\mathbf{p}_t-\mathbf{q}_t\right)\mathbf{n}_{E_t}dt
\end{equation}
Here $k_{ee}$ is a scalar stiffness constant. Time intervals $[t_a^i,t_b^i]\in[0,1]$ are the penetration time intervals, defined as time while the two edges are penetrating each other and $i$ is the number of penetration time intervals. $\mathbf{n}_t$, $\mathbf{p}_t$ and $\mathbf{q}_t$ denote contact normal, position of the contact point on two edges during $\Delta t$. The collision point on the two edges can again be expressed using barycentric coordinates of two vertices of the edge as $\displaystyle \mathbf{p}_t = w_a\mathbf{a}_t+w_b\mathbf{b}_t$ and $\displaystyle \mathbf{q}_t = w_c\mathbf{c}_t+w_d\mathbf{d}_t$. The penalty force $\mathbf{I}_p^{EE}$ is then applied to the object mesh $\displaystyle \left(-\mathbf{I}_p^{EE} \text{for pulling} \right)$. A force of the same magnitude but opposite direction is applied to the haptic proxy.

\subsection{Local deformation of the mesh: clay-like behaviour}
While deforming the object we want to replicate a clay-like behaviour in our model, i.e., the object should be malleable near the point where external force is applied but the movement of the whole structure of the object should be negligible due to this external force. To this end, we define a kernel function $G_d$ in Equation~\ref{EQ:kernel_damp} around the position of haptic proxy. The velocities of the object mesh are scaled with the weights of the kernel. If $\displaystyle r = ||\mathbf{x}-\mathbf{x}_c||_2$, then, 
\begin{equation}\label{EQ:kernel_damp}
    \displaystyle G_{d}(\mathbf{x}) =
    \begin{cases}
     \displaystyle \frac{1}{1+k_1 r} & \text{if} \;\;\; r < R_D \\
     \displaystyle \frac{1}{1+k_1 r+\exp{(k_2 r)}} & \text{if} \;\;\; r \geq R_D
     \end{cases}
\end{equation}
where $k_1$, $k_2$ are stiffness constants, $\mathbf{x}_c$ denotes the position of haptic proxy and $||\cdot||_2$ denotes the $l_2$ norm. $R_D$ is the influence radius of the damping kernel. As a result, velocities of points further away from the haptic proxy are more damped than closer points.

\subsection{Haptic Rendering: Wetting a Mesh}
Haptic force in wetting tool is rendered using the same technique we employ for deforming a mesh. But unlike deformation, the mesh geometry remains unchanged.

\section{Multi-Resolution, Multi-Timescale Simulation Framework}\label{sec:multires}

Physics-based simulation is computationally expensive and cannot be performed on extremely high resolution meshes at interactive rates. However, visual fidelity suffers a lot when low resolution meshes are used. On the other hand, haptic fidelity requires simulations to run at very high frame rates. Our framework allows us to find common ground between all these disparate goals.
\begin{figure}
    \centering
    \includegraphics[width = 0.9\columnwidth]{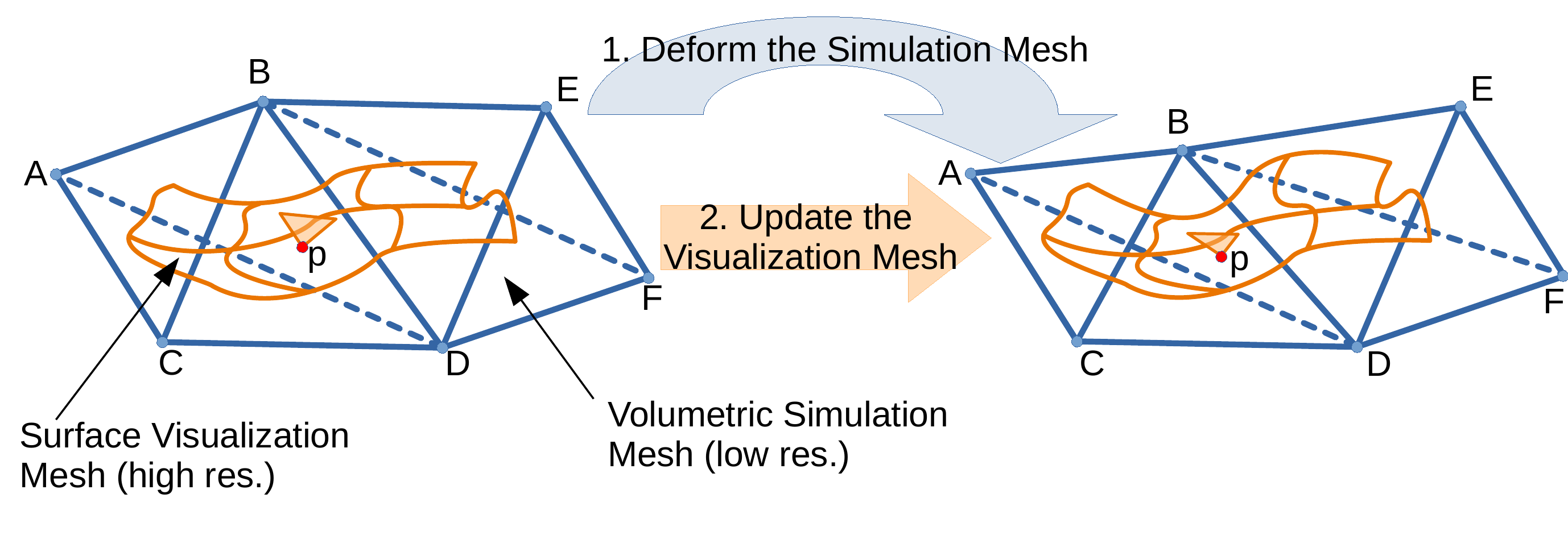}
    \caption{Multi-resolution, multi-timescale deformation}
    \label{fig:multires-deform}
\end{figure}

\begin{figure}
    \centering
    \includegraphics[width = 0.9\columnwidth]{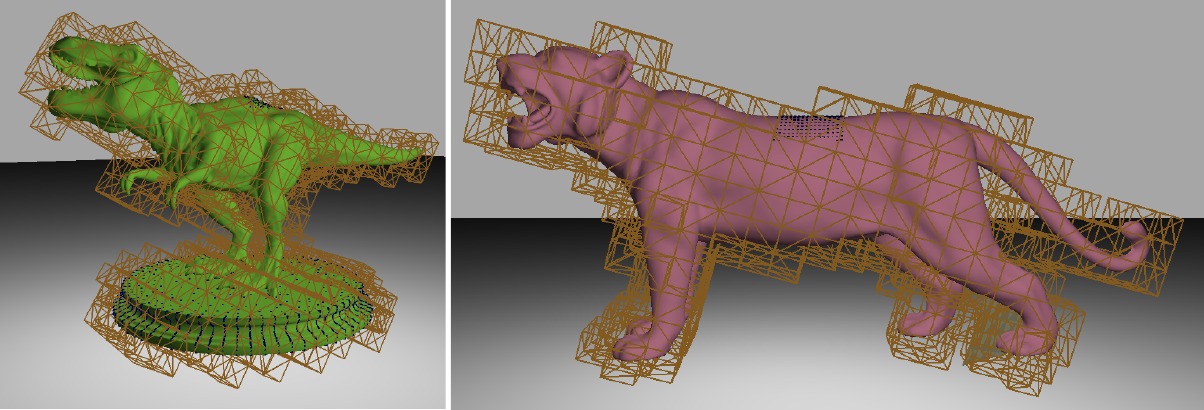}
    \caption{Surface visualization mesh embedded inside volumetric simulation mesh: T-Rex (left) and Pink-Panther (right).}
    \label{fig:sec_cage}
\end{figure}

Our simulation runs on a coarse, low resolution volumetric mesh with tetrahedral elements that encloses a high resolution surface mesh with triangle elements like a cage. As shown in Figure~\ref{fig:multires-deform}, the vertices of the surface mesh are expressed in the local space of the simulation mesh using barycentric coordinates. When the simulation mesh is deformed, the barycentric coordinates of the surface mesh vertices in the local space of the simulation mesh do not change. This lets us calculate new coordinates of the surface mesh vertices in a global coordinate system. Similar ideas can be found in~\cite{cage_1}~\cite{cage_2}. In Figure~\ref{fig:sec_cage} two high resolution surface meshes, T-Rex (left) \& Pink-Panther (right), and their corresponding low resolution volumetric simulation meshes are depicted. 
Any manipulation performed on the simulation mesh gets transferred to the visualization surface mesh using a weight kernel. This sets up the multi-resolution component of our framework.

\subsection{Transfer of Deformation}
\begin{figure}
    \centering
    \includegraphics[width=0.9\columnwidth]{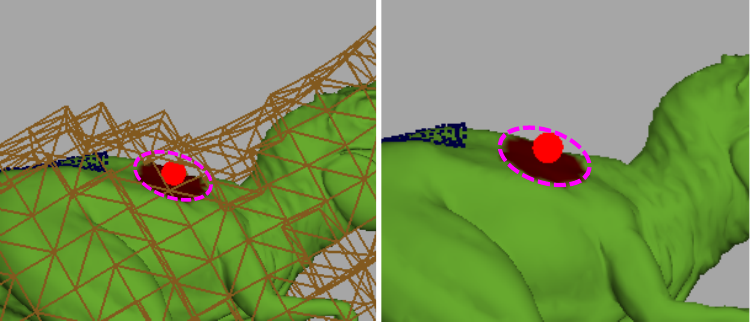}
    \caption{A deformation tool is colliding with the simulation mesh (left). The circled color gradient indicates the region of deformation on the visualization mesh. The deformation is projected onto the surface mesh for visualization (right).}
    \label{fig:tool_cage_push}
\end{figure}
When an deformation (push/pull) tool collides with the outer surface of the tetrahedral simulation mesh, we visualize it by projecting a region with color gradient on the surface mesh to denote the deformation region (Figure~\ref{fig:tool_cage_push} left). The deformations are performed on the simulation mesh. Using barycentric coordinates this deformation is then transferred to the surface mesh for visualization (Figure~\ref{fig:tool_cage_push} right).

\subsection{Transfer of Wetting}
During wetting any node of secondary mesh gets the same saturation value of the tetrahedron which contains it.

\subsection{Multi-Timescale Haptic and Visual Feedback}
For smooth haptic force feedback a minimum of refresh rate of $\displaystyle 1000$ frames/sec is required. On the other hand for smooth visual feedback a refresh rate of $\displaystyle 60$ frames/sec is sufficient. To achieve both these requirements, the whole simulation is run in two distinct threads. We use MS Windows's in-built WINAPI package to create them. On one thread, physical simulations along with graphics rendering are performed while other thread is used for rendering haptic feedback. We have kept the haptic thread running at $\displaystyle 1000$ frames/sec all time using HAPI API which samples the haptic force feedback at required rate. Refresh rate of the visual thread varies from $\displaystyle 70-900$ frames/sec, depending on the underlying object mesh. So, the haptic thread keeps rendering the same old force feedback at the higher frame rate, until it gets a force update from the visual thread which runs at a much slower rate. The synchronization between the two threads is obtained implicitly due to rapid update rate of the haptic thread, instead of using a blocking, explicit synchronization construct. This is common practice followed by many haptics literature~\cite{dis_pen_1}\cite{ccd_hap_forc}. Because of this construct, our framework can work on high resolution mesh models with intricate details without degrading the quality of a user's tactile experience.  

\section{Results}\label{result}
In this section we present results that help evaluate the performance of our mesh reshaping solution. First, we show the results of deforming the object mesh. We then demonstrate the effect of wetting the material. Further, we present the results of a user study, conducted to evaluate the qualitative performance of our solution. A quantitative evaluation of our framework is also conducted to affirm that we satisfy real-time interaction constraints.

All the results presented here are obtained on a system with a Intel i$7$-$4770$K octa-core processor at $3.5$GHz, $32$GB RAM, a single Nvidia Geforce GTX Titan GPU with $5860$ MB of graphics memory and a $6$-DOF haptic device from Geomagic Touch.
\subsection{Deforming and Wetting the Object Mesh}
\subsubsection{Deforming}
\begin{figure}[h!tb]
    \centering
    \includegraphics[width=0.95\columnwidth]{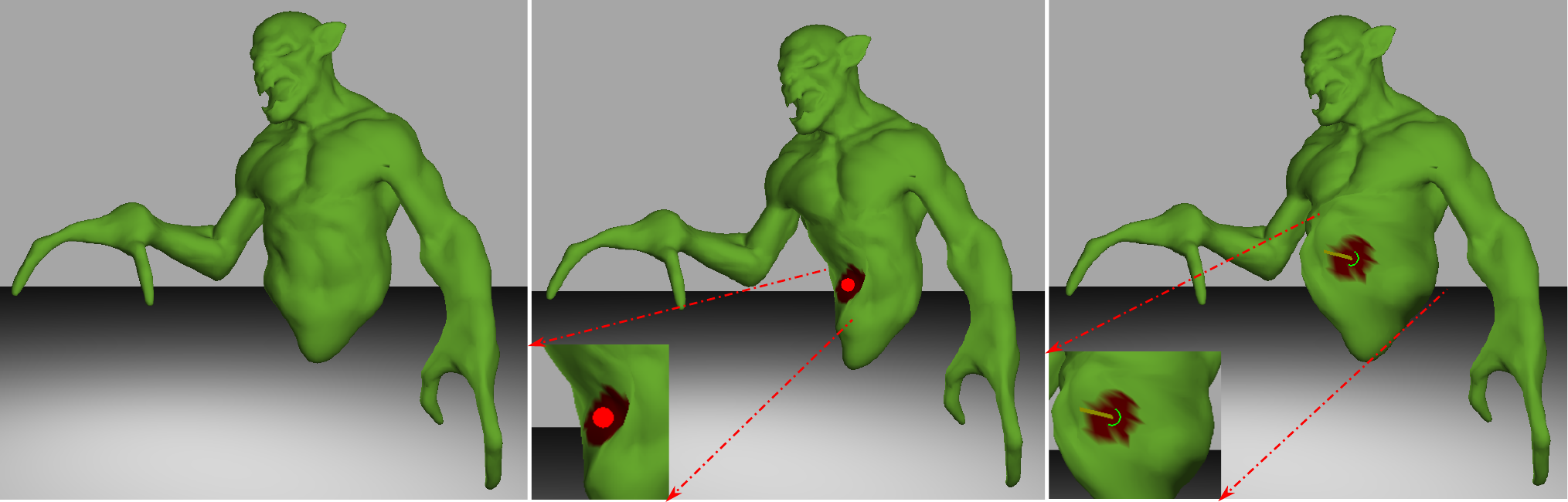}
    \caption{Illustration of original model (left), interaction of push tool with the model (middle) and  interaction of pull tool with the model (right).}
    \label{fig:press_all}
\end{figure}

As shown in Figures~\ref{fig:press_all}, whenever a haptic proxy touches the simulation mesh, a color gradient gets projected on the surface visualization mesh near proxy within radius $R_D$ as mentioned in Equation~\ref{EQ:kernel_damp}. This helps the user to get a better perception of deformation region. In Figure~\ref{fig:press_all}, a zombie object mesh with a push deformation (middle) and a pull deformation (right) are shown. For push deformation the mesh collapses and bulging takes place for pull deformation.

\subsubsection{Wetting}
\begin{figure}[h!tb]
    \centering
    \includegraphics[width=0.9\columnwidth]{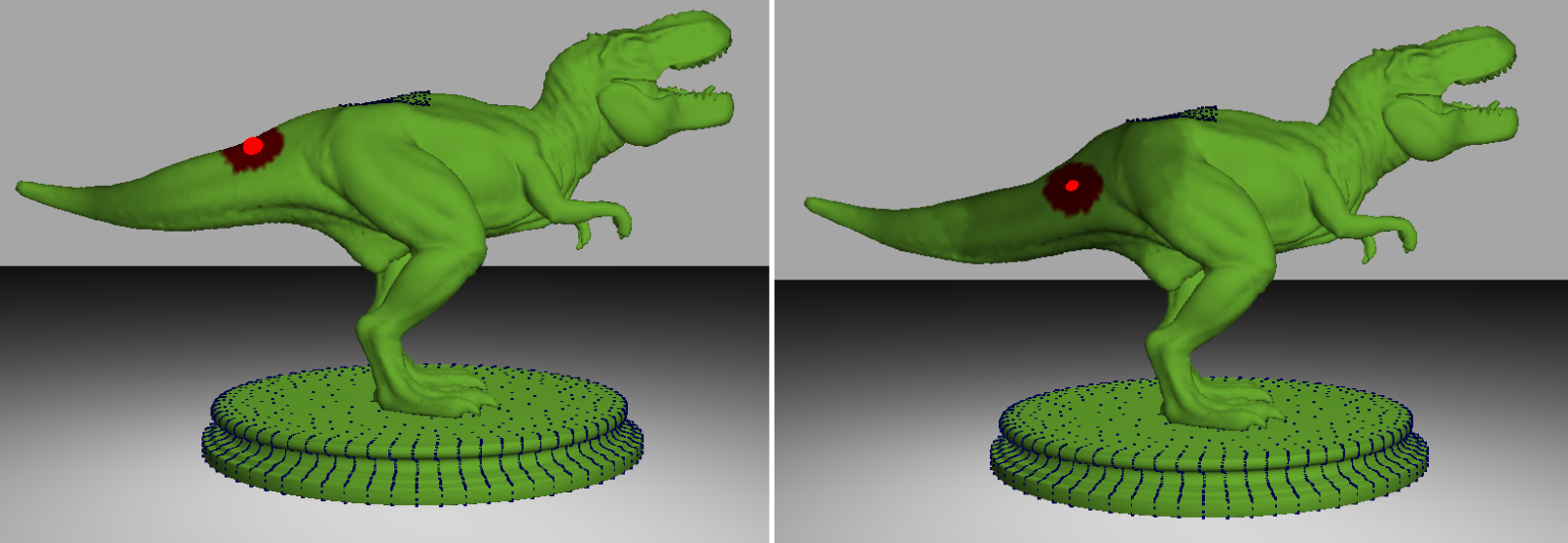}
    \caption{Deforming a dry (left) and partially wet (right) T-Rex model with a push haptic tool.}
    \label{fig:wet_dry_trex}
\end{figure}

\begin{figure}[h!tb]
    \centering
    \includegraphics[width=0.9\columnwidth]{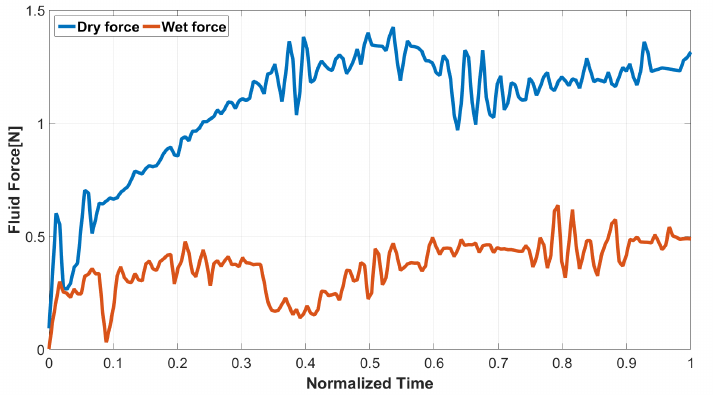}
    \caption{Illustration of haptic feedback force while interacting with dry and wet object. As expected the force is much less for the wet case.}
    \label{fig:wet_dry_forc}
\end{figure}

Using a wetting tool we can wet a material by transfer of fluid. Any vertex of surface mesh gets the same saturation value of the tetrahedron from the simulation mesh that contains the vertex. In Figure~\ref{fig:wet_dry_trex} a user is shown interacting with a dry and a partially wet T-Rex model. Except the effect of wetting on elasticity, the other material properties of the object mesh and the area where the user interacts remain same in both the cases. The  haptic  feedback  during  this  interaction  is  shown in Figure~\ref{fig:wet_dry_forc}. The perceptive change of haptic force feedback after fluid absorption is evident from the plot which indicates that the wet model offers less resistance compared to dry one. Moreover, as shown in Figure~\ref{fig:wet_dry_trex}, the wet portion of the mesh exhibits more deformation due to change of material property after water absorption.

\begin{figure}[h!tb]
    \centering
    \includegraphics[width=0.9\columnwidth]{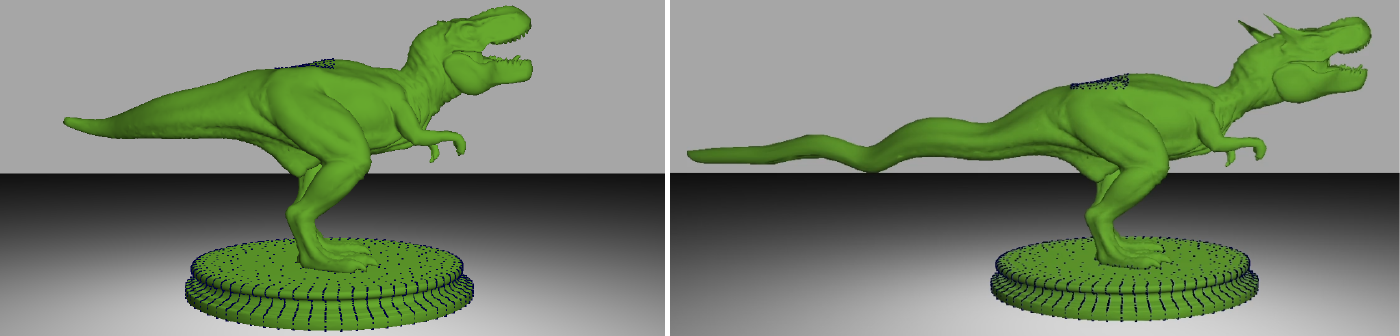}
    \caption{Original (left) and deformed (right) T-Rex model.}
    \label{fig:sculpt_trex}
\end{figure}

In Figure \ref{fig:sculpt_trex}, we present a T-Rex model reshaped using our framework.

\subsection{User Study}
A user study was conducted to evaluate the subjective quality of our virtual mesh reshaping solution compared to real world experience.

\subsubsection{Study Subjects}
$20$ subjects in the age group $20-35$ years participated in the user study. All the participants confirmed that they are not differently-abled either physically or mentally. None of the participants had any prior experience with a haptic setup.

\subsubsection{Study Setup}
As the subjects participating in our experiment were not familiar with any kind of haptics setup, we first trained them to use a haptic device. For that purpose, we used a model scene provided with Geomagic Touch haptic device. The scene contains two boxes and using a haptic proxy, a user can move or lift those boxes while getting appropriate force feedback. We ask each of the subjects to move and lift the boxes with the haptic proxy repeatedly until he/she feels comfortable handling a haptic device.

\subsubsection{Study Steps}
\begin{figure}[h!tb]
    \centering
    \includegraphics[width=0.9\columnwidth]{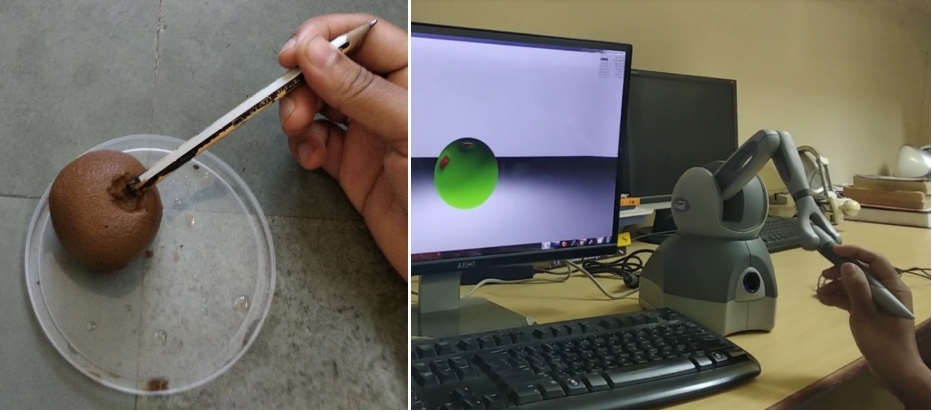}
    \caption{Comparing the experience of deforming a real clay sphere on the left to the haptic feedback of deforming a virtual clay sphere on the right.}
    \label{fig:userstudy}
\end{figure}

We perform our evaluation based on a double stimulus comparison~\cite{sub_test} method. The steps of the evaluation are as pointed out below.
\begin{itemize}
    \item \textbf{First stimulus}: We ask the subjects to mould a ball of clay to any shape of their choice using their hands and a pencil to get a feel of real world shape shaping.
    \item \textbf{Second stimulus}: The subjects are then asked to reshape object models virtually using our framework with and without haptic feedback.
\end{itemize}
Our experimental setup is shown in Figure~\ref{fig:userstudy}. A user deforming a real clay sphere (left) and a virtual clay sphere (right) is shown in the figure. After the experiment is finished, the participants are asked to rate their experience on a scale of $1$ (very poor) to $5$ (very good) for the following parameters.
\begin{itemize}
    \item \textbf{Haptic on/off}: The users are asked to reshape any mesh with and without haptic feedback, and then considering without feedback as baseline, rate how much their interactivity experience improves with haptic feedback.
    \item \textbf{Visual-haptic synchronization}: The participants are asked if they experienced any delay between visual change and haptic force feedback.
    \item \textbf{Realistic}: The users are asked to rate how close is their experience compared to real world.
    \item \textbf{Physical consistency}: Consistency of the visual simulation of our framework with real world physical objects.
\end{itemize}

The mean and standard deviation of the scores of the user feedback opinions are listed in Table~\ref{tab:table_user}. The ratings reflect highly realistic experience with little difference of opinion (low standard deviation).

\begin{table}
  \centering
  \caption{Mean and standard deviation of user feedback}~\label{tab:table_user}
  \begin{tabular}{|l|r|r|}
    \hline
        \textbf{\textit{Parameter}}
        & \textbf{\textit{Mean}}
        & \textbf{\textit{Std}} \\
    \hline
    \hline
        Haptic on/off & 4.70 & 0.28 \\
        Visual-haptic sync & 4.72 & 0.25 \\
        Realistic & 4.61 & 0.35 \\
        Physical consistency & 4.48 & 0.29 \\
    \hline
  \end{tabular}
\end{table}

\subsection{Quantitative Evaluation}
As mentioned earlier, in our multi-timescale framework, the haptic thread updates at $\displaystyle 1000$ frames per second for smooth interaction. Depending on the model structure used, the frame rate of visual rendering thread varies between $\displaystyle 70$ and $\displaystyle 900$ frames per second which is sufficient for smooth visual feedback. Further, to speed up interaction frame rate in the graphics thread we parallelized the computations on the GPU wherever possible, using Nvidia CUDA. The interactive frame rates of different tools of our framework are presented in Table~\ref{tab:table_quantity}.

\begin{table}[h!tb]
  \centering
  \caption{Average frame-rate for different tools.}~\label{tab:table_quantity}
  \begin{tabular}{|l|r|r|}
    \hline
        \textbf{\textit{Model}}
        & \textbf{\textit{Push/Pull}}
        & \textbf{\textit{Wet/Dry}} \\
    \hline
    \hline
        T-Rex & 71.3 & 792.6\\
        Zombie & 87.9 & 863.7\\
        Sphere/Cylinder & 85.9 & 841.1\\
    \hline
  \end{tabular}
\end{table}

\section{Conclusion and Future Work}\label{conclusion}
We present a novel approach for stable multi-resolution, multi-timescale simulation framework for mesh reshaping, enhanced with haptic feedback and physically accurate material simulation. We devise solutions to numerous challenges like wetting of materials and the consequent simulation of variable elasticity, and deformable porous solid simulation. Finally, we evaluate the appeal and interactivity of our solution via a user study and a variety of simulation results. One of the major limitations of our work is that it works only with one initial mesh. There is no provision for adding more meshes, and cannot thus model the functionality of material deposition. In future, we want to work in this direction.

\bibliographystyle{IEEEtran}
\bibliography{IEEEabrv, bibliography}

\end{document}